\documentclass[3p,12pt]{elsarticle}
\usepackage[colorlinks=true,citecolor=blue,linkcolor=blue]{hyperref}
\usepackage{amsmath}
\usepackage{amssymb}
\usepackage{color}
\usepackage{graphicx}
\usepackage{longtable}
\usepackage{listings}
\usepackage{chngcntr}
\counterwithin{figure}{section}
\usepackage[dvipsnames]{xcolor}
\usepackage{todonotes}
\usepackage{array, booktabs, makecell}
\newcounter{bla}

\DeclareFixedFont{\ttb}{T1}{txtt}{bx}{n}{10} % for bold
\DeclareFixedFont{\ttm}{T1}{txtt}{m}{n}{10}  % for normal
\DeclareFixedFont{\tti}{T1}{txtt}{it}{n}{10}  % for normal

\usepackage{color}
\definecolor{deepblue}{rgb}{0,0,0.5}
\definecolor{deepred}{rgb}{0.6,0,0}
\definecolor{deepgreen}{rgb}{0,0.5,0}
\usepackage{url}

\usepackage{listings}

\usepackage{titlesec}
\titleformat{\paragraph}[runin]
{\bfseries\scshape}{\theparagraph}{1em}{}

\newcommand\pythonstyle{\lstset{
language=Python,
basicstyle=\ttm,
otherkeywords={self},             % Add keywords here
keywordstyle=\ttb\color{deepblue},
emph={MyClass,__init__, uls},          % Custom highlighting
emphstyle=\ttb\color{deepred},    % Custom highlighting style
stringstyle=\color{deepgreen},
frame=tb,                         % Any extra options here
showstringspaces=false,            % 
commentstyle=\tti,
morecomment=[s]{"""}{"""},
}}
\newcommand\bashstyle{\lstset{
language=bash,
basicstyle=\ttm,
otherkeywords={self},             % Add keywords here
keywordstyle=\ttb\color{deepblue},
emph={uls-calc,uls-scan,uls-nest},          % Custom highlighting
emphstyle=\ttb\color{deepred},    % Custom highlighting style
stringstyle=\color{deepgreen},
frame=tb,                         % Any extra options here
showstringspaces=true,            % 
commentstyle=\tti,
}}

\lstnewenvironment{python}[1][]
{
\pythonstyle
\lstset{#1}
}
{}

\lstnewenvironment{bash}[1][]
{
\bashstyle
\lstset{#1}
}
{}

\newcommand\pythoninline[1]{{\pythonstyle\lstinline!#1!}}

\usepackage{allrunes}
\usepackage{epiolmec}
\usepackage[dvipsnames]{xcolor}

\usepackage{lineno}
\usepackage{listings}
\usepackage{eso-pic}
\usepackage{epsfig,graphicx}
\usepackage{bm} %letras griegas en negrita (uso: \bm{\alpha})
\usepackage{slashed}
\newcommand*\cpp{C\kern-0.2ex\raisebox{0.4ex}{\scalebox{0.8}{+\kern-0.4ex+}}}

\usepackage{color}
\usepackage{pstricks}

\usepackage{fancyhdr}
\usepackage{textcomp}

\usepackage{xcolor} 
\usepackage{colortbl} 
 
\newcommand{\unit}{\leavevmode\hbox{\small1\kern-3.6pt\normalsize1}}

\def \GeV{{\mathrm{GeV}}}

\newcommand{\equaref}[1]{Eq~(\ref{#1})}

\newcommand{\figref}[1]{Fig~(\ref{#1})}

\newcommand{\secref}[1]{Section~(\ref{#1})}
\newcommand{\secrefs}[2]{Sections~(\ref{#1}-\ref{#2})}

\newcommand{\tabref}[1]{Table~(\ref{#1})}

\newcommand{\listing}[1]{Listing~(\ref{#1})}
\def \ULYSSES{\sc ULYSSES}
\usepackage{pgf}
\begin{document}
\AddToShipoutPictureBG*{%
  \AtPageUpperLeft{%
    \hspace{0.95\paperwidth}%
    \raisebox{-\baselineskip}{%
      \makebox[1pt][r]{FERMILAB-PUB-20-275-T, SISSA 17/2020/FISI, IPPP/20/30}
}}}%

\bibliographystyle{unsrt}
\begin{frontmatter}
\title{{\ULYSSES}: Universal LeptogeneSiS Equation Solver}

\author[a]{A.~Granelli}
\author[b]{K.~Moffat}
\author[c,d,e]{Y. F.~Perez-Gonzalez}
\author[f]{H.~Schulz}
\author[c]{J.~Turner}

\address[a]{SISSA/INFN, Via Bonomea 265, I-34136 Trieste, Italy.}
\address[b]{Institute for Particle Physics Phenomenology, Durham University, Durham, UK}
\address[c]{Fermi National Accelerator Laboratory, Batavia, IL, 60510-0500, USA}
\address[d]{Department of Physics \& Astronomy, Northwestern University, Evanston, IL 60208, USA}
\address[e]{Colegio de F\'isica Fundamental e Interdisciplinaria de las Am\'ericas (COFI), 254 Norzagaray street, San Juan, Puerto Rico 00901}
\address[f]{Department of Physics, University of Cincinnati, Cincinnati, OH 45219, USA}

\journal{Computer Physics Communications}

\begin{abstract}
{\ULYSSES} is a python package that calculates the baryon asymmetry 
produced from leptogenesis in the context of a type-I seesaw mechanism. The
code solves the semi-classical Boltzmann  equations for points in the model
parameter space as specified by the user. We provide  a selection of
predefined Boltzmann equations as well as a plugin mechanism for externally provided models
of leptogenesis. Furthermore, the {\ULYSSES} code provides tools for
multi-dimensional parameter space exploration. The emphasis of the code is on
user flexibility  and rapid evaluation. It is publicly available at \url{https://github.com/earlyuniverse/ulysses}.
\end{abstract}
\end{frontmatter}

\section{Introduction}\label{sec:overview}
The  two leading theories that explain  the excess of matter over antimatter are leptogenesis \cite{Fukugita:1986hr} and electroweak baryogenesis \cite{Shaposhnikov:1987tw,Cohen:1990it}.
The latter theory has attracted much attention given its close relation with  Higgs physics and much of the 
model parameter space has been explored. 
The former, in its various manifestations, appeals to many given its  connection to neutrino masses and mixing.
Although the  mechanisms which generate the baryon asymmetry in both scenarios are vastly different, a common
feature  is the need to solve Boltzmann equations (BE) for points in the relevant model parameter space.
{\ULYSSES} is a \emph{python} package that solves the semi-classical BE for leptogenesis in the context of a type-I seesaw mechanism
and, to the authors knowledge, is the first publicly available code for this task.

The provided momentum-averaged BEs 
are based on the out-of-equilibrium decays of right-handed neutrinos and resonant 
leptogenesis. 
Effects such as lepton flavour, scatterings and spectator processes are also provided if the user wishes to apply them.  
For a given point in the model parameter space, {\ULYSSES} calculates the final baryon asymmetry (provided in terms of the baryon-to-photon ratio, $\eta_{B}$, the baryonic yield, $Y_B$, and the baryonic density parameter,  $\Omega_B h^2$) and plots the  lepton asymmetry number density as a function of 
the evolution parameter. 
For the user who wishes to undertake a multi-dimensional exploration of the parameter space,  we provide instructions on how to use {\sc Multinest} \cite{Feroz:2008xx} in combination with  {\ULYSSES}. This 
allows visualisation of the multi-dimensional parameter space which is consistent with the measured baryon-to-photon ratio \cite{Patrignani:2016xqp,Ade:2015xua}.
We have designed the code in a modular fashion, separating the physics of the baryon asymmetry production
from the parameter space exploration. 

As this paper is a manual on how to use the {\ULYSSES} code, we refrain from discussing the different regimes and subtleties of 
the leptogenesis mechanism and instead refer the reader to Refs. (\cite{Hagedorn:2017wjy,Dev:2017trv,Dev:2017wwc}) for broad reviews on various aspects of thermal and resonant leptogenesis.

The paper is organised as follows: in \secref{sec:conventions} we  discuss the parametrisation and normalisation conventions {\ULYSSES} applies.
In \secref{sec:plugins}, we describe the preprovided BEs and follow in \secref{sec:install} with  installation instructions and a  discussion of code dependencies. 
In \secref{sec:code}, we explain the structure of the code and show the user how to calculate the baryon asymmetry for a  point  in the model parameter space. 
Scripts and examples of multi-dimensional parameter space exploration, as well as user options, 
 are presented in 
\secref{sec:runtime}  and  finally we make concluding remarks in \secref{sec:conclusions}.

\section{Conventions}\label{sec:conventions}
We begin in \secref{sec:yukmat} by providing details on our parametrisation of the Yukawa matrix  and then follow
in \secref{sec:normalisation} with a discussion of our applied normalisation of the BEs.

\subsection{Yukawa matrix}\label{sec:yukmat}
 One of  the simplest extensions of the Standard Model (SM) that explains small neutrino masses is the type-I seesaw mechanism \cite{Minkowski:1977sc,Yanagida:1979as,GellMann:1980vs}.
Leptogenesis can be regarded as a cosmological consequence of the seesaw mechanism and provides an elegant way of explaining tiny
neutrino masses and the baryon asymmetry of the Universe.

 This mechanism introduces a set of heavy right-handed Majorana neutrino fields $N_i$ and augments the SM Lagrangian to include the following terms
\begin{equation}
\mathcal{L} = i\overline{N_{i}}\slashed{\partial}N_{i}  -Y_{\alpha i}\overline{L_{\alpha}}\tilde{\Phi}N_{i}-\frac{1}{2}M_{i}\overline{N^c_{i}}N_{i} + \text{h.c.}\,,
\end{equation}
where $Y$ is the Yukawa matrix and $\Phi$ the Higgs  doublet, $\Phi^T = \left( \phi^+, \phi^0 \right)$ and $\tilde{\Phi} = i \sigma_2 \Phi^*$, and $L^T = \left( \nu^T_L, l^T_L  \right)$ is the leptonic doublet. For convenience and without loss of generality, we have chosen the basis in which the Majorana mass term is diagonal. 
After electroweak symmetry breaking, at  tree-level, the light neutrino mass matrix (at first order in the seesaw expansion) is
\begin{equation}\label{eq:treemass}
m^{\text{tree}} \approx m^{}_D M^{-1} m_D^T\,,
\end{equation}
where  $m_D = Y v$ is the Dirac mass matrix that develops once the Higgs acquires the vacuum expectation value, $v$. We  use the conventions that
 $v=174.0$ GeV and  $m^{\text{tree}}$ does not have a minus sign. We parametrise the Yukawa matrix in analogy with Casas and Ibarra \cite{Casas:2001sr}:
\begin{equation}\label{eq:CandI}
Y=\frac{1}{v}U\sqrt{\hat{{m}}_{\nu}}R^T\sqrt{M_R}\,,
\end{equation}
 where $U$ is the leptonic mixing matrix, $\hat{{m}}_{\nu}$ is the  diagonal light neutrino mass matrix, $R$ is 
a complex, orthogonal matrix and $M_R$ is the diagonal mass matrix of the heavy right-handed neutrinos.
Using this parametrisation the  model parameter space is 18 dimensional where nine parameter are associated to the low-energy
scale physics  and the remaining nine
parameters are associated to the high-scale physics of the right-handed neutrinos. This parametrisation has the benefit that neutrino masses and mixing from oscillation data are
recovered\footnote{While the Casas-Ibarra parametrisation is convenient and widely used, {\ULYSSES} allows
the user to provide their own Yukawa matrix and we detail how to do this in \secref{sec:code}.}.

We apply the PDG convention  \cite{Patrignani:2016xqp} to parametrise the PMNS matrix:
\begin{equation}
\begin{aligned}
U =&\begin{pmatrix}
1 & 0 & 0 \\
0 & c_{23} & s_{23}  \\
0 &- s_{23} & c_{23} 
\end{pmatrix}
\begin{pmatrix}
c_{13} & 0 & s_{13}e^{-i\delta} \\
0 & 1 & 0 \\
-s_{13}e^{i\delta} &0 & c_{13}
\end{pmatrix}
\begin{pmatrix}
c_{12} & s_{12} & 0\\
-s_{12} & c_{12} & 0\\
0 & 0 & 1
\end{pmatrix}
\begin{pmatrix}
1 & 0 & 0\\
0&e^{i\frac{\alpha_{21}}{2}} & 0\\
0 & 0 &  e^{i\frac{\alpha_{31}}{2}}
\end{pmatrix}\,,
\end{aligned}
\end{equation}
 where $c_{ij} \equiv \cos\theta_{ij}$, $s_{ij} \equiv \sin\theta_{ij}$, $\delta$ is the Dirac  phase and $\alpha_{21}$, $\alpha_{31}$ are the Majorana phases
 which vary between $0 \leq\alpha_{21}, \alpha_{31} \leq 4\pi$.
The $R$-matrix has the form:
\begin{equation}
R=\begin{pmatrix}
1 & 0 & 0 \\
0 & c_{\omega_{1}} & s_{\omega_{1}} \\
0 &- s_{\omega_{1}} & c_{\omega_{1}} 
\end{pmatrix}
\begin{pmatrix}
c_{\omega_{2}} & 0 & s_{\omega_{2}} \\
0 & 1 & 0\\
-s_{\omega_{2}} & 0 & c_{\omega_{2}} 
\end{pmatrix}\\
\begin{pmatrix}
c_{\omega_{3}} & s_{\omega_{3}} & 0\\
-s_{\omega_{3}} & c_{\omega_{3}} & 0\\
0 & 0 & 1
\end{pmatrix}\,,
\end{equation}
where $c_{\omega_{i}} \equiv \cos\omega_{i}$, $s_{\omega_{i}} \equiv \sin\omega_{i}$ 
 and the complex angles are given by $\omega_{i} \equiv x_{i}+iy_{i}$ for $x, y$  free, real parameters.
 
 The above parametrisation does not account for the radiative corrections to
 the light neutrino masses from the $Z$, $W$  and Higgs boson. In some regions
 of the parameter space these corrections can be sizeable, such that the
 tree and one-loop contributions to the mass are comparable in magnitude
 \cite{LopezPavon:2012zg}. As the tree and one-loop level contributions enter
 with different signs, a small neutrino mass compatible with data may be the
 consequence of cancellation between these two contributions. Such fine-tuning
 was quantified in \cite{Moffat:2018wke} and depending on the user specified fine-tuning
 tolerance\footnote{One can check that the two-loop contribution
 to the light neutrino mass is not larger than the one-loop contribution. This
 procedure is outlined in \cite{Moffat:2018wke}.}, the  correction to the
 Casas-Ibarra parametrisation can be implemented in {\ULYSSES} using \cite{Lopez-Pavon:2015cga}
  \begin{equation}\label{eq:CIloop}
Y=\frac{1}{v}U\sqrt{\hat{m}_{\nu}}R^T\sqrt{f(M)^{-1}}\,,
\end{equation}
where 
\begin{equation}
m_{\nu} = m^{\text{tree}}+m^{\text{1-loop}}\,,
\end{equation}
with 
\begin{equation}
\begin{aligned}
& m^{\text{1-loop}} = \\
& - m^{}_D \left( \frac{M}{32 \pi^2 v^2} \left(\frac{\log\left(\frac{M^2}{m_H^2}\right)}{\frac{M^2}{m_H^2}-1} + 3 \frac{\log\left(\frac{M^2}{m_Z^2}\right)}{\frac{M^2}{m_Z^2}-1}\right) \right) m_D^T\\
& = - \frac{1}{32 \pi^2 v^2} m_D \text{diag} \left(g\left(M_{1}\right), g\left(M_{2}\right), g\left(M_{3}\right) \right) m_D^T\,,
\end{aligned}
\end{equation}
and
\begin{equation}
g\left(M_i\right) \equiv M_i \left(\frac{\log\left(\frac{M_i^2}{m_H^2}\right)}{\frac{M_i^2}{m_H^2}-1} + 3 \frac{\log\left(\frac{M_i^2}{m_Z^2}\right)}{\frac{M_i^2}{m_Z^2}-1}\right)\,.
\end{equation}
The contribution from two-loop corrections is usually small as these will be suppressed by an extra factor of the Yukawa couplings squared and a further factor $\mathcal{O}(10^{-2})$ from the loop integral. 
The matrix $m_{\nu}$ is rewritten in the factorised form using the leptonic mixing matrix:
\begin{equation}
m_{\nu}  = U \hat{m}_{\nu} U^T\,,
%\hat{m}_{\nu} = U^{\dagger} m_{\nu} U^*\,,
\end{equation}
where $\hat{m}_{\nu}$ is the positive diagonal matrix of light neutrino masses. The inclusion of the loop effect  is a command line
argument that we detail in  \secref{sec:code}.

\subsection{Normalisation and conversion of lepton to baryon asymmetry}\label{sec:normalisation}
The baryon asymmetry may be parametrised by the baryon-to-photon ratio, $\eta_B$, which is 
defined to be
\begin{equation}
\eta_B \equiv \frac{n_B-n_{\overline{B}}}{n_\gamma}\,,
\end{equation}
where $n_B$, $n_{\overline{B}}$ and $n_\gamma$ are the number densities of baryons, anti-baryons and photons, respectively.
This quantity can be measured using two independent methods that
probe the Universe at different stages of its evolution. Big-Bang nucleosynthesis (BBN)  \cite{Patrignani:2016xqp} and 
Cosmic Microwave Background radiation (CMB) data  \cite{Ade:2015xua} are given by
\begin{equation}
\begin{aligned}
{\eta_{B}}_{\text{BBN}}  & = \left(5.80-6.60\right)\times 10^{-10}\,, \\
{\eta_{B}} _{\text{CMB}}& = \left(6.02-6.18\right)\times 10^{-10}\,,
\end{aligned}
\end{equation}
at 95$\%$ CL, respectively. As the uncertainties of the CMB measurement are smaller than those from BBN, this is the
value taken in the code for the MultiNest scans. For completeness, {\ULYSSES} also returns the baryonic yield and baryonic density parameter which follow from the baryon-to-photon ratio:
\begin{equation}
%\begin{aligned}
Y_{B}   = \eta_B\cdot\frac{45 \zeta(3)}{\pi^4g_{*,s}(t_{\rm rec})}\,,\quad\quad \Omega_Bh^2  = \eta_B\cdot\frac{m_pn_\gamma}{\rho_ch^{-2}}\,,
%\end{aligned}
\end{equation}
where $g_{*,s}(t_{\rm rec})=43/11$ are the entropic effective degrees of freedom at present, $m_p$ is the proton mass and $\rho_c$ is the critical density of the Universe~\cite{Patrignani:2016xqp}.

The ULYSSES code solves BEs in terms of number densities of particles, or particle
asymmetries, normalised to a comoving volume which contains one photon.
This is equivalent to choosing the normalised equilibrium abundance of the right-handed neutrino to be $N_N^{eq}\left(z\right) =3/8 \cdot z^2K_{2}(z)$
which is the  same convention applied in \cite{Buchmuller:2004nz}.
Therefore, the conversion from the $B-L$ number density to the baryon-to-photon ratio is as follows:
\begin{equation}
\eta_B \equiv \frac{N_{B}}{N^{\rm rec}_{\gamma}} = a_{\rm sph} \frac{N_{B-L}}{N^{\rm rec}_{\gamma}} = \frac{28}{79}\frac{1}{27} N_{B-L} = 0.013 N_{B-L}\,,
\end{equation}
where $N_{B-L}$ is the final $B-L$ asymmetry, $ a_{\rm sph} = 28/79$ is the Standard model sphaleron factor and the $1/27$ factor derives from 
the dilution of the baryon asymmetry by photons for our choice of normalisation\footnote{We note that another common convention is to normalise to one ultrarelativistic right-handed neutrino per comoving volume, see for example  Ref.~\cite{Marzola:2012xwq}.}. New physics can change the sphaleron factor, for instance in the
supersymmetric Standard Model, $a_{\rm sph} = 8/23$. This will alter the overall normalisation factor, referred to as ``normfact'', which multiplies $N_{B-L}$. To allow for such new physics, normfact can be altered by the user through a command line option  as detailed in \secref{sec:commonopts}.

\section{Built-in Boltzmann equations}\label{sec:plugins}
In this section, we list and briefly discuss the preprovided BEs that are shipped with
 {\ULYSSES}. We refer to BEs that incorporate off-diagonal flavour oscillations as density matrix
 equations (DME). 
  The density matrix equations solved  can be found in Ref.~\cite{Blanchet:2011xq} while in the resonant case we solve the equations of
Ref.~\cite{DeSimone:2007edo}. Finally, the model which includes scattering is
based on Ref.~\cite{Buchmuller:2004nz}. We provide example parameter cards for each
model. They are located in the \path{examples} folder of the source tree.
A quick overview of the contents of this section can be found in~\tabref{tab:models}.
The information about currently available models is also accessible by invoking the
command {\texttt{uls-models}} which is available after installation of {\ULYSSES}.
We note that for all of the preprovided BEs, we have assumed  a standard
cosmology. From this assumption, the Boltzmann equations can be
    written in terms of the scale factor $a$ 
    which can be converted to an evolution in time, $t$. The time variable can
    be exchanged for a more convenient evolution parameter $z=M/T$ where $M$ is
    the mass of the light right-handed neutrino and $T$ is the plasma temperature. If the
    Hubble expansion rate evolved according to standard cosmology this is a
    convenient approach. We provide one example (1BE1Fsf) where 
    the Hubble expansion rate is explicit and the evolution parameter is the scale factor.
    This would be a starting point for the user interested in implementing their own
    non-standard cosmology.
    
\begin{table}[t!]
    \centering
    \begin{tabular}{l|l|l}
        \toprule
       Model & example input file & Description  \\ 
        \midrule
        1DME     & 1N3F.dat & DME 1 RHN \\
        2DME     & 2N3F.dat & DME 2 RHN \\
        3DME     & 3N3F.dat & DME 3 RHN \\
        \midrule
        1BE1F    & 1N1F.dat & one-flavoured BE  1  RHN \\
        1BE2F    & 1N2F.dat & two-flavoured BE  1 RHN \\
        1BE3F    & 1N3F.dat & three-flavoured BE 1  RHN\\
        \midrule
        2BE1F    & 2N1F.dat & one-flavoured BE with 2  RHN \\
        2BE2F    & 2N2F.dat & two-flavoured BE with 2  RHN \\
        2BE3F    & 2N3F.dat & three-flavoured BE with 2  RHN \\
        \midrule
        3DMEsct & 3N3F.dat & DME  3  RHN including scattering effects \\
        1BE1Fsf &  1N1F.dat & 1BE1F  evolving in scale factor \\
%        3DMErhtau && \cite{Blanchet:2008hg} \\
2RES & Res.dat & 2BE3F in the resonant regime\\
2RESsp & Res.dat &  2RES including spectator processes \\
        \bottomrule 
    \end{tabular}
    \caption{Overview of built-in plugins. We abbreviate density matrix equations, Boltzmann equations and (decaying) right-handed neutrino as DME, BE and RHN respectively. The evolution variable is $z=M_{1}/T$ for all plugins other than 1BE1Fsf which evolves in the cosmological scale factor.}
    \label{tab:models}
\end{table}

\begin{itemize}
\item  \textbf{1DME}  provides the semi-classical density matrix equations (DME) for one 
decaying right-handed neutrino: \begin{equation}\label{eq:etab3DME}\begin{aligned}
\frac{d N_{N_{1}}}{d z} &=-D_{1}\left(N_{N_{1}}-N_{N_{1}}^{\mathrm{eq}}\right) \\
\frac{d N_{\alpha \beta}^{B-L}}{d z} &=\epsilon_{\alpha \beta}^{(1)} D_{1}\left(N_{N_{1}}-N_{N_{1}}^{\mathrm{eq}}\right)-\frac{1}{2} W_{1}\left\{P^{0(1)}, N^{B-L}\right\}_{\alpha \beta}\\
&-\frac{\Gamma_{\tau}}{2 H z}\left[\left(\begin{array}{ccc}
1 & 0 & 0 \\
0 & 0 & 0 \\
0 & 0 & 0
\end{array}\right),\left[\left(\begin{array}{ccc}
1 & 0 & 0 \\
0 & 0 & 0 \\
0 & 0 & 0
\end{array}\right), N^{B-L}\right]\right]_{\alpha \beta}\\
&-\frac{\Gamma_{\mu}}{2 H z}\left[\left(\begin{array}{ccc}
0 & 0 & 0 \\
0 & 1 & 0 \\
0 & 0 & 0
\end{array}\right),\left[\left(\begin{array}{ccc}
0 & 0 & 0 \\
0 & 1 & 0 \\
0 & 0 & 0
\end{array}\right), N^{B-L}\right]\right]_{\alpha \beta}\,,
\end{aligned}\end{equation}
where  $N_{B-L}$, $D_1$ and $W_1$ are the
    (negative) lepton asymmetry number density, decay and washout
    respectively. This equation accounts for the  transitions between the 1, 2 and 3-flavour
regimes by promoting the lepton asymmetry number density to a 
density matrix and adding the appropriate commutators for flavour
effects involving the interaction widths, $\Gamma$, of the leptons.  The
initial conditions for RH neutrino and lepton asymmetry number densities are
set to zero initial abundance; however, this can be easily modified by the
user.

\item  \textbf{2DME}  provides the DMEs  for the decay
 of two heavy neutrinos. This is the same as \equaref{eq:etab3DME} but with
 subscript $1$ replaced with a dummy index $i$ that is summed over two heavy
 mass states.

\item  \textbf{3DME}  provides the DMEs  for the decay of three heavy
    neutrinos.

\item \textbf{1BE1F}  provides the semi-classical BE for one decaying
    right-handed neutrino, $N_1$, with number density $N_{N_1}$ in the single
    flavour approximation.  The BE is given by 
    \begin{equation}\begin{aligned} \frac{d N_{N_{1}}}{d z}
    &=-D_{1}\left(N_{N_{1}}-N_{N_{1}}^{\mathrm{eq}}\right)   \\ \frac{d
    N_{B-L}}{d z} &=\epsilon^{(1)}
    D_{1}\left(N_{N_{1}}-N_{N_{1}}^{\mathrm{eq}}\right)-W_{1} N_{B-L}\,,
    \end{aligned}\end{equation} 
    This is the simplest possible Boltzmann equation for thermal
    leptogenesis.

 \item \textbf{1BE2F}  provides the semi-classical two-flavoured BE for one decaying right-handed
neutrino with flavour effects due to tau leptons.  The kinetic equations solved are
\begin{equation}\label{eq:1BE2F}\begin{aligned} &\frac{d N_{N_{1}}}{d
        z}=-D_{1}\left(N_{N_{1}}-N_{N_{1}}^{\mathrm{eq}}\right)\\ &\frac{d
        N_{\alpha \alpha}}{d z}=\sum_{\alpha = \beta
        \tau}\left(\epsilon_{\alpha \alpha}^{(1)}
        D_{1}\left(N_{N_{1}}-N_{N_{1}}^{\mathrm{eq}}\right)-p_{1 \alpha} W_{1}
    N_{\alpha \alpha}\right)\,, \end{aligned}\end{equation}
    where $p_{1 \alpha}$ are projection probabilities between the mass and flavour
    states. The state $\beta$ is the coherent $e/\mu$ superposition that is left
    after $\tau$ decoheres.

\item \textbf{1BE3F} provides the semi-classical three-flavoured BE for one
    decaying right-handed neutrino. This BE is accurate for $M_1 \lesssim
    10^{9}$ GeV and the differential equations solved are
    \begin{equation}\begin{aligned} &\frac{d N_{N_{1}}}{d
            z}=-D_{1}\left(N_{N_{1}}-N_{N_{1}}^{\mathrm{eq}}\right)\\ &\frac{d
            N_{\alpha \alpha}}{d z}=\sum_{\alpha = e, \mu,
            \tau}\left(\epsilon_{\alpha \alpha}^{(1)}
            D_{1}\left(N_{N_{1}}-N_{N_{1}}^{\mathrm{eq}}\right)-p_{1 \alpha}
        W_{1} N_{\alpha \alpha}\right)\,, \end{aligned}\end{equation}
    where $p_{1 \alpha}$ are projection probabilities between the mass and
    flavour states, computed from the $c_{i \alpha}$ elements in lines 14 to
    16.

 \item \textbf{2BE1F} provides the semi-classical BE for two decaying
     right-handed neutrinos in the single flavour approximation.
       The solved
     equations are
    \begin{equation}\label{eq:eta2BE}
    \begin{aligned} \frac{d N_{N_{i}}}{d z}
            &=-D_{i}\left(N_{N_{i}}-N_{N_{i}}^{\mathrm{eq}}\right) \\ \frac{d
            N_{B-L}}{d z} &=\sum_{i=1}^{2} \left( \epsilon^{(i)}
                D_{i}\left(N_{N_{i}}-N_{N_{i}}^{\mathrm{eq}}\right)-W_{i}
    N_{B-L}\right)\,,
      \end{aligned}\end{equation}
    where $i\in\{1,2\}$.

 \item \textbf{2BE2F} provides the semi-classical BE for two decaying
     right-handed neutrinos in the two-flavour approximation.

\item \textbf{2BE3F} provides the semi-classical three-flavoured BE for two
    decaying right-handed neutrinos.

\item  \textbf{3DMEsct}  provides the three heavy neutrino density
    matrix equations including $\Delta L = 1$ scattering effects. These are the
    same as \equaref{eq:etab3DME} but with three heavy neutrinos and
    the replacement \begin{equation} D_1 \rightarrow D_1 + S_1\,,
    \end{equation} where $S_1$ incorporates the effects of $\Delta L = 1$
    scatterings involving $N_1$.

\item  \textbf{1BE1Fsf} is based on the same set of Boltzmann
    equations as 1BE1F but rather than evolving in $z=M/T$ evolves in the
    scale factor. 
    This BE is useful if the user wants to implement
    a  non-standard cosmology which modifies the Hubble expansion rate which
    is given explicitly in this code (for examples see
    Refs.~\cite{Dutta:2018zkg,Buchmuller:2011mw}). We note that in this BE we use the normalisation
    convention of \secref{sec:normalisation}, namely the particle number density is normalised to a comoving volume 
    which contains a single photon.
    
\item  \textbf{2RES} provides two heavy neutrino Boltzmann equations for
    the resonant case. These are the same equations as for 2BE3F,
    however the CP asymmetries are modified for accuracy in resonant scenarios
    in which $M_2 - M_1 \sim \Gamma_i$. The modified CP asymmetries used are~\cite{DeSimone:2007edo}
    \begin{equation} \begin{aligned} -\epsilon_{\alpha \alpha}^{(i)}&=\sum_{j
            \neq i} \frac{\operatorname{Im}\left[Y_{i \alpha}^{\dagger}
                Y_{\alpha j}\left(Y^{\dagger} Y\right)_{i
            j}\right]+\frac{M_{i}}{M_{j}} \operatorname{Im}\left[Y_{i
                \alpha}^{\dagger} Y_{\alpha j}\left(Y^{\dagger} Y\right)_{j
        i}\right]}{\left(Y^{\dagger} Y\right)_{i i}\left(Y^{\dagger}
        Y\right)_{j j}}\left(f_{i j}^{\operatorname{mix}}+f_{i
        j}^{\mathrm{osc}}\right),\\ f_{i
        j}^{\operatorname{mix}}&=\frac{\left(M_{i}^{2}-M_{j}^{2}\right) M_{i}
        \Gamma_{j}}{\left(M_{i}^{2}-M_{j}^{2}\right)^{2}+M_{i}^{2}
        \Gamma_{j}^{2}},\quad f_{i
        j}^{\mathrm{osc}}=\frac{\left(M_{i}^{2}-M_{j}^{2}\right) M_{i}
        \Gamma_{j}}{\left(M_{i}^{2}-M_{j}^{2}\right)^{2}+\left(M_{i}
                \Gamma_{i}+M_{j} \Gamma_{j}\right)^{2}
    \frac{\operatorname{det}\left[\operatorname{Re}\left(Y^{\dagger}
    Y\right)\right]}{\left(Y^{\dagger} Y\right)_{i i}\left(Y^{\dagger}
    Y\right)_{j j}}}\,.  \end{aligned} \end{equation}

\item  \textbf{2RESsp} provides the equations for resonant
    leptogenesis with the lowest temperature scale spectator effects included
    through the factors $C^{\Phi}$ and $C^l$ \cite{Nardi:2006fx} by promoting
    the washout terms to
    \begin{equation}
    - p_{1 \alpha} W_1 \sum_{\beta} \left( C^l_{\alpha \beta} +
    C^{\Phi}_{\beta} \right) N_{\beta \beta}\,.
    \end{equation}
    The current implementation includes spectator effects accurate for $T \ll 10^{8}$ GeV.
\end{itemize}

\section{Installation}\label{sec:install}

The code is hosted on \url{https://github.com/earlyuniverse/ulysses}. 
Once the git repository is pulled,
the  basic installation steps are shown in \listing{lst:install}.  In
addition, releases are packaged and available to install with {\sc pip} from
\url{pypi.org}.

\begin{minipage}{\linewidth}
\lstset{label={lst:install}}
\lstset{caption={Minimal installation steps.}}
\begin{bash}
# Installation from within the source tree
git clone https://github.com/earlyuniverse/ulysses.git
cd ulysses
pip install . --user

# Installation with pip or pip3 from pypi.org
pip install ulysses --user
\end{bash}
\end{minipage}

\subsection{Core dependencies}
The code is written in python3 and heavily uses the widely available modules {\sc
NumPy}~\cite{oliphant2006guide,5725236} and {\sc SciPy}~\cite{2020SciPy-NMeth}.
We accelerate the computation with the just in time compiler provided by {\sc
Numba}~\cite{10.1145/2833157.2833162} where meaningful.  At its core,
{\ULYSSES} solves a set of coupled differential equations. To undertake this
task we use {\sc odeintw} \cite{odeintw} which provides a wrapper of
scipy.integrate.odeint that allows it to handle complex and matrix differential
equations. The latter is redistributed with {\ULYSSES} and does not need to be
downloaded separately.  These dependencies for {\ULYSSES} are automatically
resolved during the install process with pip. They provide the minimal
functionality for solving Boltzmann equations for a given point in the model
parameter space. 

\subsection{Additional requirements for multidimensional scans}

\begin{minipage}{\linewidth}
\lstset{label={lst:installmultinest}}
\lstset{caption={Installation of libMultiNest}}
\begin{bash}
git clone https://github.com/JohannesBuchner/MultiNest
cd MultiNest/build
cmake ..
make
cd ..
export LD_LIBRARY_PATH=$PWD/lib:$LD_LIBRARY_PATH
\end{bash}
\end{minipage}

For multidimensional parameter space exploration with the aim of finding
regions compatible with the experimentally measured $\eta_B$ we provide a
script, {\texttt uls-nest}, which invokes
MultiNest~\cite{2014A&A...564A.125B,Feroz:2013hea,Feroz:2008xx}.
MultiNest efficiently scans a parameter space to find regions of maximum likelihood. 
{\sc uls-nest} implements a simple log-likelihood for that purpose:

\begin{equation}\label{eq:llog}
%\begin{aligned}
    \log\mathcal{L}(x|\vec{p}) =  -0.5 \cdot \left(\frac{\eta_B(\vec{p}) - x}{\Delta x}\right)^2, 
%\end{aligned}
\end{equation}
where $x\pm\Delta x$ are the experimentally measured values ${\eta_{B}} _{\text{CMB}} = (6.10\pm0.04)\times 10^{-10}$.
We denote the baryon asymmetry parameter as calculated by {\ULYSSES} for a point $\vec{p}$ of the currently loaded model as $\eta_B(\vec{p})$. 

MultiNest is a code written in C and FORTRAN that can optionally be compiled with
support for message passing interface (MPI) to enable parallel computing on a single workstation
or potentially many network connected computers. The usage of MultiNest in python is made
possible with the additional pip installable package {\sc pymultinest}
(\url{https://github.com/JohannesBuchner/MultiNest}). {\sc pymultinest}
requires a shared library of MultiNest to be be available in the users
environment. MPI parallelism is available through {\sc
mpi4py}~\cite{DALCIN20051108,DALCIN2008655,DALCIN20111124}. It should be noted
that {\sc mpi4py} and {\sc pymultinest} are automatically installed when using
pip to install {\ULYSSES}. The compilation of the MultiNest library cannot be
automated in that fashion. An example of how to obtain the source code and how
to compile the shared library using {\sc cmake} is given in
\listing{lst:installmultinest}. Furthermore, {\sc cmake} detects if MPI is available
on the system and triggers the compilation of the library {\sc libmultinest\_mpi} in addition
to the serial {\sc libmultinest}.

\section{Computing model}\label{sec:code}

We designed {\ULYSSES} to be easily extensible in such a way that users can
focus on the physics. The module contains a single base class, 
ULSBase, which has all the infrastructure needed to solve the problem at
hand. This includes machineries to set global constants, parameters of the
physics models and the ODE solver as well as commonly used computations, such
as the calculation of the PMNS matrix in the Casas-Ibarra parametrisation.  The
base class itself is devoid of any concrete physics but contains a dummy
function, {\texttt{EtaB}}, which must be overwritten in classes which are
derived from ULSBase that implement the actual Boltzmann equations.
We further provide a plugin mechanism that allows the seamless usage of user
developed models with the run-time scripts of {\ULYSSES} --- as long as the new
model also derives from {\ttm ULSBase} and implements its own version of
{\texttt{EtaB}}. An example of the code structure can be seen in
~\listing{lst:codeskeleton}.

\begin{minipage}{\linewidth}

\lstset{caption={A skeleton for an externally provided plugin model for the calculation of $\eta_B$}}
\lstset{label={lst:codeskeleton}}
\begin{python}
# Content of myplugin.py
import ulysses
class EtaB_plugin(ulysses.ULSBase):
    """
    My new plugin
    """
    def RHS(self):
        """
        Right hand side of ODE system goes here
        """
        rhs = ... 
        return rhs

    @property
    def EtaB(self):
        """
        Invoke e.g. odeintw, calculate and return etab.
        """
        y0      = np.array([0+0j,0+0j], dtype=np.complex128)
        ys, _   = odeintw(self.RHS, y0, self.zs)
        nb      = self.normfact*(ys[-1,1]+ys[-1,2]+ys[-1,3])
        return np.real(nb) 

\end{python}
\end{minipage}

\subsection{Setting parameters}\label{sec:input}

\begin{table}[t!]
    \centering
    \begin{tabular}{llcc}
        \toprule
        Parameter                                                         & variable name      & default               & unit            \\
        \midrule
        Higgs VEV,                                     $v$                & {\texttt{vev}}      & $174.0$               & [GeV]           \\
        Higgs mass,                                    $M_H$              & {\texttt{mhiggs}}   & $125.0$               & [GeV]           \\
        Z boson mass,                                  $M_{Z}$            & {\texttt{mz}}       & $91.1876$             & [GeV]           \\
        Planck mass,                                   $M_{\text{PL}}$    & {\texttt{mplanck}}  & $1.22\times10^{19}$   & [GeV]           \\
        Neutrino cosmological mass,                    $m^*$              & {\texttt{mstar}}    & $10^{-12}$            & [GeV]           \\
        Degrees of freedom,                            $g*$               & {\texttt{gstar}}    & $106.75$              &                 \\
        Normalisation factor                                                  & {\texttt{normfact}}    & $0.013$              &                 \\
        Solar mass square splitting,                   $m^2_{\text{SOL}}$ & {\texttt{m2solar}}  & $7.4\times10^{-23}$   & [GeV$\strut^2$] \\
        Atm. mass squared splitting (normal),          $m^2_{\text{ATM}}$ & {\texttt{m2atm}}    & $2.515\times10^{-21}$ & [GeV$\strut^2$] \\
        Atm. mass squared splitting (inverted),        $m^2_{\text{ATM,inv}}$ & {\texttt{m2atminv}} & $2.483\times10^{-21}$ & [GeV$\strut^2$] \\
        \bottomrule 
    \end{tabular}
    \caption{Overview of global parameters and their defaults values.}
    \label{tab:const}
\end{table}
\begin{minipage}{.19\textwidth}%
    \begin{center}
\lstset{label={lst:examplecalc}}
\lstset{caption={Example input for uls-calc.}}
\begin{python}
m       -1.10
M1      12.10
M2      12.60
M3      13.00
delta  213.70
a21     81.60
a31    476.70
x1      90.00
x2      87.00
x3     180.00
y1    -120.00
y2       0.00
y3    -120.00
t12     33.63
t13      8.52
t23     49.58
\end{python}
    \end{center}
\end{minipage}%
\begin{minipage}{.075\textwidth}%
    \hfill
\end{minipage}%
\begin{minipage}{.28\textwidth}%
    \begin{center}
\lstset{label={lst:examplescan}}
\lstset{caption={Example input for uls-scan.}}
\begin{python}
m       -1.10
M1       6.00   12.00
M2      12.60
M3      13.00
delta  213.70
a21     81.60
a31    476.70
x1      90.00
x2      87.00
x3     180.00
y1    -120.00
y2       0.00
y3    -120.00
t12     33.63
t13      8.52
t23     49.58
\end{python}
    \end{center}
\end{minipage}%
\begin{minipage}{.075\textwidth}%
    \hfill
\end{minipage}%
\begin{minipage}{.33\textwidth}%
    \begin{center}
\lstset{label={lst:examplenest}}
\lstset{caption={Example parameter card for uls-nest.}}
\begin{python}
m       -4.00   -1.00
M1       6.00   
M2       7.00
M3       7.50
delta    0.00  360.00
a21      0.00  720.00
a31      0.00  720.00
x1       0.00  180.00
x2       0.00  180.00
x3       0.00  180.00
y1    -180.00  180.00
y2    -180.00  180.00
y3    -180.00  180.00
t12     33.63
t13      8.52
t23     49.58
\end{python}
    \end{center}
\end{minipage}%

\begin{minipage}{.19\textwidth}%
    \begin{center}
\lstset{label={lst:examplecalcfree}}
\lstset{caption={Example input for uls-calc.}}
\begin{python}
Y11_mag  0.01
Y12_mag  0.01 
Y13_mag  0.01
Y21_mag  0.01
Y22_mag  0.03
Y23_mag  0.05
Y31_mag  0.01
Y32_mag  0.03
Y33_mag  0.05
Y11_phs -1.11
Y12_phs  2.89
Y13_phs  1.32
Y21_phs  2.88
Y22_phs -0.23
Y23_phs -1.80
Y31_phs -1.72
Y32_phs  2.96
Y33_phs  1.39
M1       12.0
M2       12.5
M3       13.0
\end{python}
    \end{center}
\end{minipage}%
\begin{minipage}{.075\textwidth}%
    \hfill
\end{minipage}%
\begin{minipage}{.28\textwidth}%
    \begin{center}
\lstset{label={lst:examplescanfree}}
\lstset{caption={Example input for uls-scan.}}
\begin{python}
Y11_mag  0.01
Y12_mag  0.01 
Y13_mag  0.01
Y21_mag  0.01
Y22_mag  0.03
Y23_mag  0.05
Y31_mag  0.01
Y32_mag  0.03
Y33_mag  0.05
Y11_phs  0.00 3.14
Y12_phs  2.89
Y13_phs  1.32
Y21_phs  2.88
Y22_phs -0.23
Y23_phs -1.80
Y31_phs -1.72
Y32_phs  2.96
Y33_phs  1.39
M1      12.00
M2      12.50
M3      13.00
\end{python}
    \end{center}
\end{minipage}%
\begin{minipage}{.075\textwidth}%
    \hfill
\end{minipage}%
\begin{minipage}{.33\textwidth}%
    \begin{center}
\lstset{label={lst:examplenestfree}}
\lstset{caption={Example parameter card for uls-nest.}}
\begin{python}
Y11_mag   0.01
Y12_mag   0.01 
Y13_mag   0.01
Y21_mag   0.01
Y22_mag   0.03
Y23_mag   0.05
Y31_mag   0.01
Y32_mag   0.03
Y33_mag   0.05
Y11_phs  -3.14 3.14
Y12_phs  -3.14 3.14
Y13_phs  -3.14 3.14
Y21_phs  -3.14 3.14
Y22_phs  -3.14 3.14
Y23_phs  -3.14 3.14
Y31_phs  -3.14 3.14
Y32_phs  -3.14 3.14
Y33_phs  -3.14 3.14
M1       12.0
M2       12.5
M3       13.0
\end{python}
    \end{center}
\end{minipage}%

All global constants are defined in the \_\_init\_\_ function of the base
class. We allow the user to set their values per the standard python keyword argument
formalism using the variable names shown in the second column
of~\tabref{tab:const}.
The required input from the user is the model parameters which derive from the
Casas-Ibarra parametrisation of the Yukawa matrix, Y, as shown in \equaref{eq:CandI}.  The parameters which may
be explored by the user are shown in \tabref{tab:params}. The lightest
neutrino mass ($m$) is fixed by the user and the two heavier neutrino masses
are fixed at the best-fit values from global fit data \cite{Esteban:2018azc}
which can be changed in ulsbase.py.  In the example shown in 
\listing{lst:examplecalc}, the lightest
active neutrino mass is $m_{1}=10^{-1.1}$ eV and the right-handed neutrino masses are
$N_{1,2,3} = 10^{12.1,12.6,13}$ GeV respectively. We note that the masses of both
the light and heavy neutrinos are provided by  the exponent to base 10. 

As discussed before, the method of Casas and Ibarra  is one popular way of
parametrising the Yukawa matrix. However,
{\ULYSSES} also allows the user to provide their own Yukawa matrix, in polar
coordinates, and calculate the resultant baryon asymmetry. We note that
the user will need to independently ensure that oscillation data is satisfied.
The input logic is such that each element of the Yukawa matrix, $Y_{ij}$, is
determined by two independent parameters {\ttm Yij\_mag} and {\ttm Yij\_phs}:

\begin{equation}
\begin{aligned}
    Y_{ij}& =  {\mathtt{ Yij\_mag}}\cdot\exp\left(i~\mathtt{Yij\_phs}\right)
\end{aligned}
\end{equation}

An example parameter card is shown in  \listing{lst:examplecalcfree}.

\begin{table}[t!]
    \centering
    \begin{tabular}{lc|ll}
    \toprule
        Parameter & Unit & \multicolumn{2}{c}{Code input example} \\ \hline
        $\delta~$                       & $\left[^\circ\right]$      &delta &270 \\
        $\alpha_{21}~$                  & $\left[^\circ\right]$      &a21   & 0\\
        $\alpha_{31}~$                  & $\left[^\circ\right]$      &a31   & 0\\
        $\theta_{23}~$                  & $\left[^\circ\right]$      &t23   & 48.7\\
        $\theta_{12}~$                  & $\left[^\circ\right]$      &t12   & 33.63\\
        $\theta_{13}~$                  & $\left[^\circ\right]$      &t13   & 8.52\\ \hline
        $ x_{1}~$                       & $\left[^\circ\right]$      &x1    & 45\\
        $ y_{1}~$                       & $\left[^\circ\right]$      &y1    & 45\\
        $ x~$                           & $\left[^\circ\right]$      &x2    & 45\\
        $ y_{2}~$                       & $\left[^\circ\right]$      &y2    & 45\\
        $ x~$                           & $\left[^\circ\right]$      &x3    & 45\\
        $ y_{3}~$                       & $\left[^\circ\right]$      &y3    & 45\\ \hline
        $\log_{10}\left(m_{1/3}\right)$ & $\left[\mathrm{eV}\right]$ &m     & -0.606206\\
        $\log_{10}\left(M_1\right)$    & [$\GeV$]                    &M1    & 11\\
        $\log_{10}\left(M_2\right)$    & [$\GeV$]                    &M2    & 12\\
        $\log_{10}\left(M_3\right)$    & [$\GeV$]                    &M3    & 15\\
            \bottomrule
    \end{tabular}
    \caption{Overview of input parameters in the Casas-Ibarra parametrisation.}
    \label{tab:params}
\end{table}

%
%
%
%{\tiny
%    \begin{minipage}{.5\linewidth}
%\end{minipage}
%}
\section{Run time scripts and examples}\label{sec:runtime}

To display the preprovided BEs, as detailed in \secref{sec:plugins}, and the strings needed to load them from the command line 
the user can call:

\begin{bash}
# display list of available models
uls-models
\end{bash}
The output is similar to \tabref{tab:models}; the shorthand for the models will be printed to screen in the leftmost column.

For convenience, we ship three runtime scripts which use the {\ULYSSES} module for
the evaluation of $\eta_B$ at a single point as well as in one-dimensional and in
multi-dimensional parameter space explorations:
\begin{itemize}
    \item {\ttm uls-calc}
    \item {\ttm uls-scan}
    \item {\ttm uls-nest}
\end{itemize}
which are discussed in  \secrefs{sec:ulscalc}{sec:ulsnest} respectively.
The only mandatory argument to all of these scripts is a parameter input card.
We allow the user to apply the Casas-Ibarra parametrisation as well as specifying the
Yukawa matrix explicitly. The former has a total of 16 free parameters, while
the latter has 21.  The structure of the parameter files is slightly different
for each script and is explained below. It should be noted that we decided
against setting the physics parameters to default values. This means that in
all scripts, the full set of 16 (21) input parameters must be provided. The
physics and computational setup can further be steered with a set of command
line options and switches. 

\subsection{Common options}\label{sec:commonopts}
We  first describe the command line options that are common to all three scripts.
All scripts allow the user to set the global constants given in \tabref{tab:const} on the
command line. The syntax is always {\ttm key:value}. For example, to set the
normalisation factor to 0.015, the user would input to the command line:

\begin{bash}
# Use one of the built-in plugin
uls-calc -m 1DME examples/1N3F.dat  normfact:0.015
\end{bash}

\paragraph{Boltzmann equation selection, -m}

The command line argument  ``-m''  is used to select a Boltzmann equation. 
For the built-in BEs this can be any string as given in
\tabref{tab:models}. For the plugin system the syntax is slightly different.
The absolute or relative path to the file containing the plugin
implementation needs to be specified, together with the
name of the class. Both are separated by a colon:

\begin{bash}
# Use one of the built-in plugins
uls-calc -m 1DME examples/1N3F.dat

# Use an externally provided plugins
uls-calc -m myplugin.py:EtaB_plugin  examples/1N3F.dat
\end{bash}

\paragraph{Inverted mass ordering, loop corrections}
By default, the normal mass ordering is applied in the calculations. To explore the parameter space
in the context of an inverted mass ordering, the command line switch ``--inv'' must be added.
Similarly, to implement  loop corrections which by default are off, as detailed in \secref{sec:yukmat},  can be enabled by adding the switch ``--loop''
to the command line.
\begin{bash}
uls-calc -m 1DME examples/1N3F.dat --inv
uls-calc -m 1DME examples/1N3F.dat --loop
uls-calc -m 1DME examples/1N3F.dat --inv --loop
\end{bash}

\paragraph{Integration range, -\,-zrange}
To set up the integration range and steps, we use the following syntax:
\begin{bash}
uls-calc -m 1DME examples/1N3F.dat --zrange 0.1,50,300
\end{bash}
This example sets the integration range to be between 0.1 and 50, using 300
steps as opposed to the default of 1000 steps between 0.1 and 1000.

\subsection{uls-calc}\label{sec:ulscalc}
This code calculates and prints the baryon asymmetry parameter for a given point 
and selected BE: 
\begin{bash}
uls-calc -m 3DME  point.txt
\end{bash}

\begin{figure}[t!]
    \centering
    \includegraphics[width=.48\textwidth]{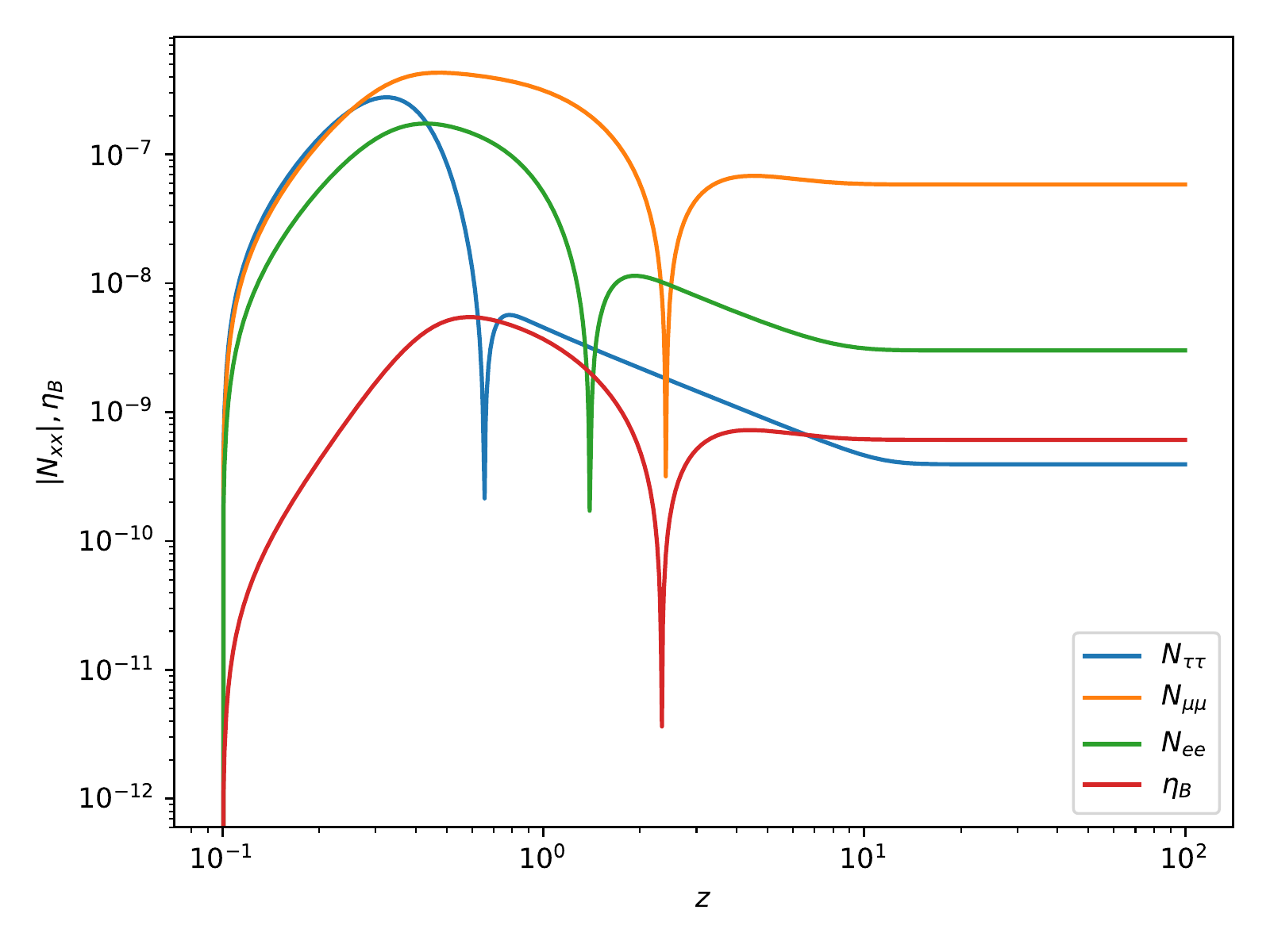}
     \includegraphics[width=.48\textwidth]{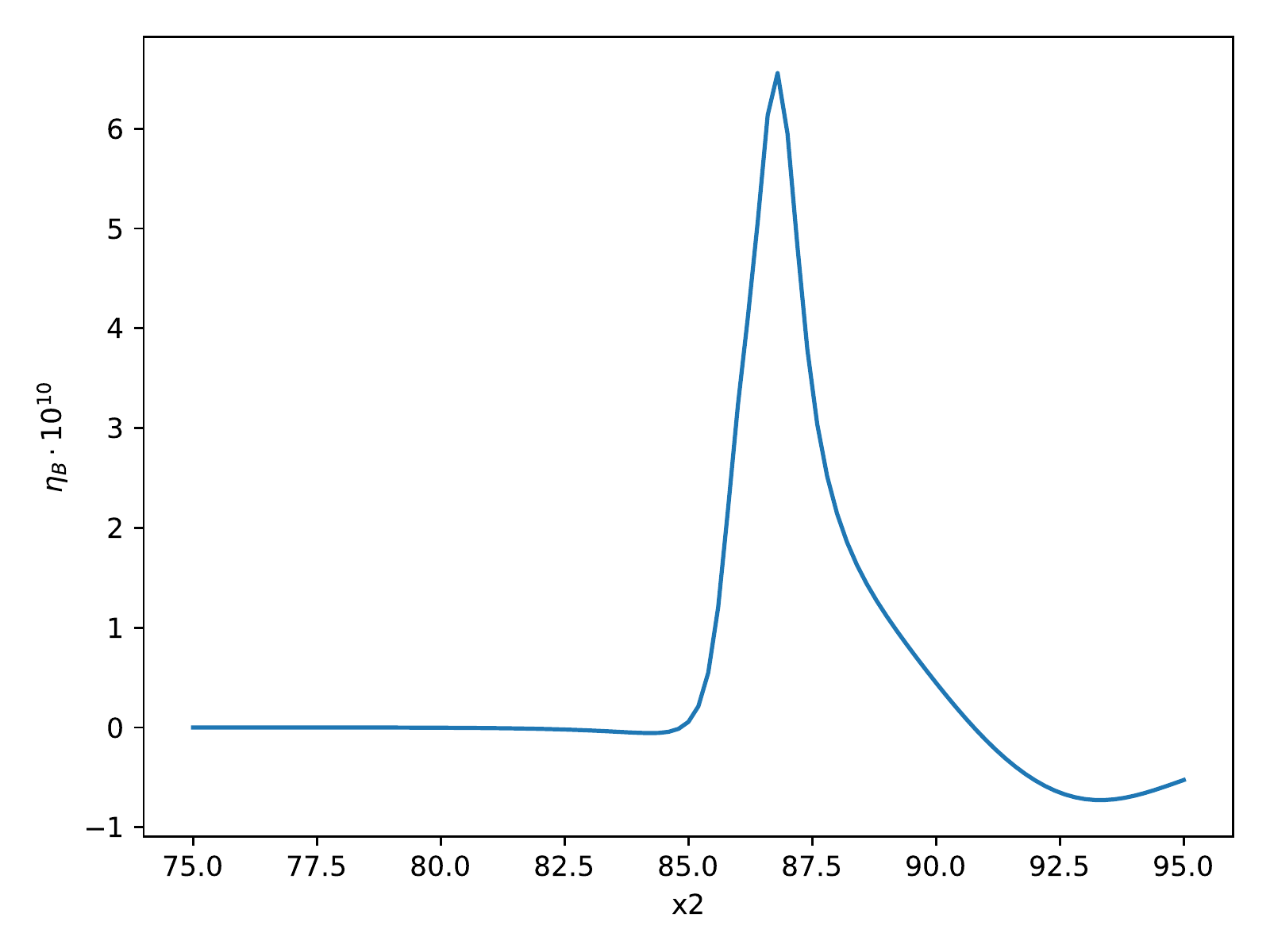}
    \caption{Example output of uls-calc (left) and uls-scan (right).}
    \label{fig:out-calc}
\end{figure}

The required positional argument is the parameter point in
question in a simple text file with parameter name value pairs. An example
parameter card is given in \listing{lst:examplecalc} for the Casas-Ibarra parametrisation
and the free format in \listing{lst:examplecalcfree}.
For convenience, we provide the functionality to write out the evolution
of the lepton asymmetry number densities if the command line option ``-o'' is provided. 
Depending on the ending of the file name this is either in the form
of a plot (see left plot of \figref{fig:out-calc}) or as an array of numbers stored in a text file.
\begin{bash}
# Produce a plot of the evolution
uls-calc -m 3DME  examples/3N3NF.dat -o evolution.pdf

# Write evolution data to a text file
uls-calc -m 3DME  examples/3N3NF.dat -o evolution.txt
\end{bash}

\subsection{uls-scan}
To perform a one-dimensional scan of $\eta_B$ for a certain model, uls-scan can
be used.  We again use the command line option ``-o'' to specify the output
file name.  An example plot of the output from uls-scan is shown in the right plot of \figref{fig:out-calc}.  The range of the parameter to be
scanned is taken from the input file (see \listing{lst:examplescan} and (\ref{lst:examplescanfree}) for an
example).  The number of points to run the scan for can be selected with
``-n'':
\begin{bash}
uls-scan -m 3DME  scan_x2.txt -o scan_x2.pdf  -n 40
\end{bash}

\subsection{uls-nest}\label{sec:ulsscan}
\begin{figure}[t!]
    \centering
    \includegraphics[width=.48\textwidth]{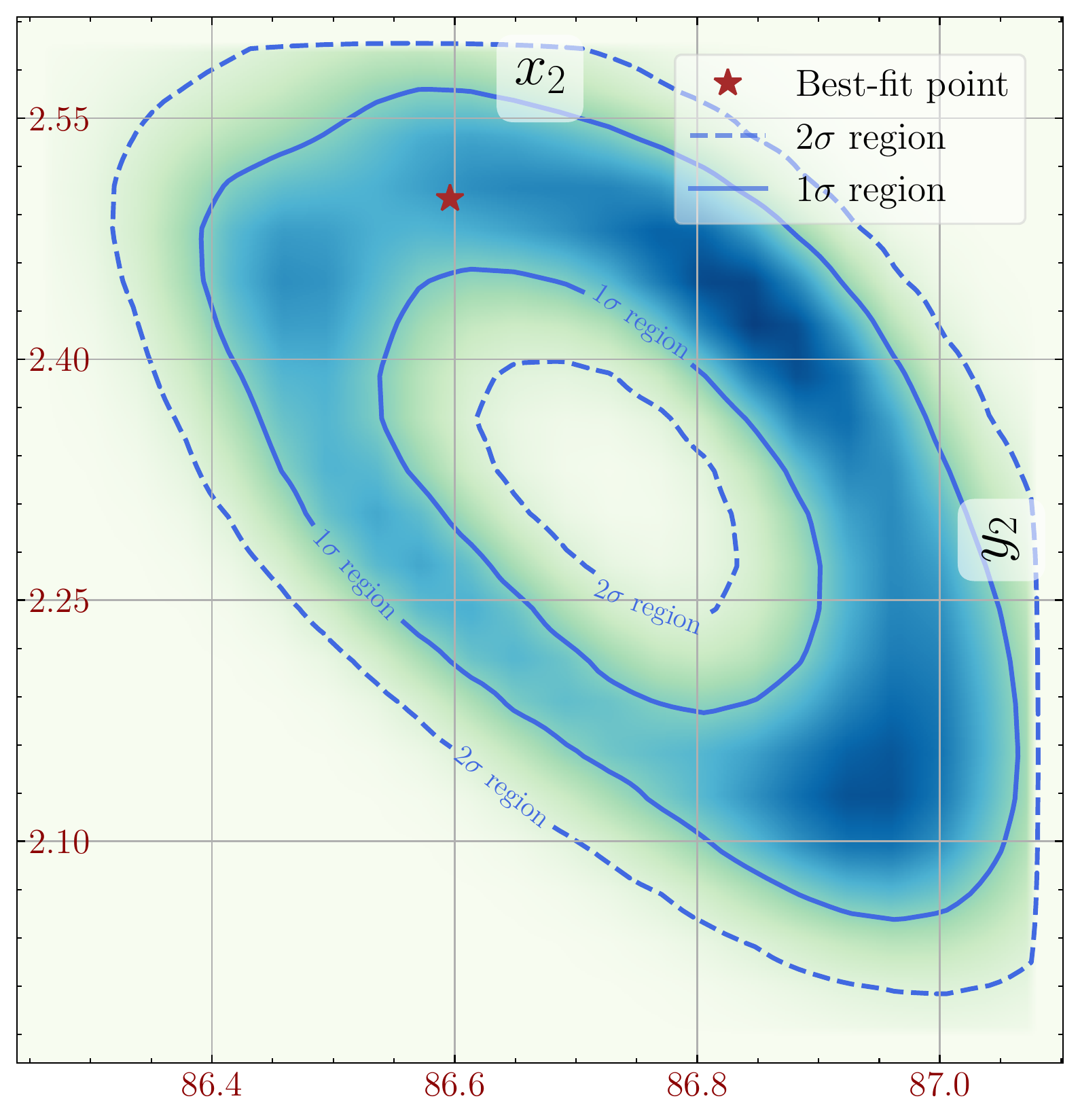}
    \caption{Visualisation of uls-nest output with SuperPlot.}
    \label{fig:out-nest}
\end{figure}
uls-nest is a multidimensional likelihood sampler using MultiNest. 

The output of uls-nest is the standard output of MultiNest which is a text file
that contains the sampled points and corresponding likelihood and posterior values.
Visualisation of the output can, for instance, be undertaken with
SuperPlot~\cite{Fowlie:2016hew} (see~\figref{fig:out-nest}) or the plotting tools that are provided by 
{\sc pymultinest}.  The parameter space to scan can be defined by supplying a simple text file
with key value pairs. We use the following logic: A parameter name followed by two numbers is
interpreted as boundaries on that particular parameter's subspace while
a single number is interpreted as fixing the corresponding parameter to
the supplied value. An example can be found in ~\listing{lst:examplenest} and (\ref{lst:examplenestfree}). 

The command line to run the code on a single CPU may look like this:
\begin{bash}
# Single core run
uls-nest -m 3DME scan_x2_y2.ranges -o 2Dscan
\end{bash}

As the computational cost increases with the number of free parameters in the
scan the run-time may become quite large. If MultiNest is compiled with MPI
enabled and {\ttm mpi4py} is installed, uls-nest can be executed in parallel. We
note that the parallel computation is already beneficial on a workstation or laptop.

\begin{bash}
# The same physics setup but distributed over 256 CPUs
mpiexec -np 256 uls-nest -m 3DME scan_x2_y2.ranges -o 2Dscan
\end{bash}

\paragraph{MultiNest parameters}\label{sec:ulsnest}
We provide access to all commonly used MultiNest parameters through command line options.
To separate them from the rest of the options, we use the pattern {\ttfamily --mn-OPTION}.
\tabref{tab:mnestpars} gives an overview of various switches and their defaults. For a thorough
discussion of their meaning we direct the reader to the official documentation at \url{https://johannesbuchner.github.io/PyMultiNest/}

\begin{table}[t!]
    \centering
    \begin{tabular}{l|c|l}
      \toprule
        Option & Default & Parameter name in pymultinest \\
        \midrule
        {\ttfamily --mn-points} & 400 & {\ttm n\_live\_points} \\
        {\ttfamily --mn-tol} & 0.5 & {\ttm evidence\_tolerance} \\
        {\ttfamily --mn-eff} & 0.8 & {\ttm sampling\_efficiency} \\
        {\ttfamily --mn-imax} & 0 & {\ttm max\_iter} \\ 
        {\ttfamily --mn-resume} & False & {\ttm resume} \\
        {\ttfamily --mn-multimodal} & False & {\ttm multimodal} \\
        {\ttfamily --mn-no-importance} & False & {\ttm not importance\_nested\_sampling} \\
        {\ttfamily --mn-seed} & -1 & {\ttm seed} \\
        {\ttfamily --mn-update} & 1000 & {\ttm n\_iter\_before\_update}\\
           \bottomrule 
    \end{tabular}
    \caption{MultiNest specific parameters and their defaults available in {\ULYSSES}. The third column identifies the parameter name as used in pymultinest.}
    \label{tab:mnestpars}
\end{table}

\section{Summary and Discussion}\label{sec:conclusions}
 {\ULYSSES} is the first publicly available code to calculate the baryon asymmetry 
in the framework of a type-I seesaw mechanism. Currently the code provides momentum-averaged Boltzmann equations for the  out-of-equilibrium decays and resonant leptogenesis with examples on how to   incorporate lepton flavour, scatterings  and spectator effects.
The  {\ULYSSES} code structure also allows the user to calculate the baryon asymmetry from their own externally
defined plugin. 
Additional effects, which would refine the baryon asymmetry calculation,  are of interest for future code development. These include
thermal production rates at finite temperature \cite{Garbrecht:2013gd,Ghisoiu:2014ena}, next-to-leading-order
corrections for the source term \cite{Bodeker:2017deo,Biondini:2016arl,Biondini:2015gyw}
and inclusion of partially equilibrated spectator processes \cite{Garbrecht:2014kda,Garbrecht:2019zaa}. Furthermore, 
inclusion of a plugin for leptogenesis via oscillation \cite{Akhmedov:1998qx} is of interest given its close connection with a number of experimental probes. Finally, we view this as a community project and invite users to add their own plugins to share with others.
This is implemented via issues and pull requests on our GitHub repository. 

\vspace{1cm}
\textbf{Acknowledgements}\\
We are deeply grateful to Serguey T. Petcov for useful discussions and suggestions. 
It is a pleasure to thank Marco Drewes for helpful discussions on this code. 
This research was supported by the Fermi National Accelerator Laboratory 
(Fermilab), a U.S. Department of Energy, Office of Science, HEP User Facility.
K.M.  acknowledges the (partial) support from the European Research
Council under the European Union Seventh Framework Programme (FP/2007-2013) / ERC
Grant NuMass agreement n. [617143].
Fermilab is managed by Fermi Research Alliance, LLC (FRA), acting under 
Contract No. DE--AC02--07CH11359. This material is based upon work supported by the U.S. Department of Energy, Office of Science, Office of Advanced Scientific
Computing Research, Scientific Discovery through Advanced Computing (SciDAC) program, grant HEP Data Analytics on HPC,
No. 1013935. It was supported by the U.S. Department of Energy
under contracts DE-AC02-76SF00515.

\bibliography{leptobib}

\end{document}